\documentclass{article}
\usepackage{amsfonts,amsmath,amscd}
\usepackage{graphicx}

\def\SO{\mathsf{SO}}
\def\so{\mathsf{so}}

\def\cD{{\cal D}}
\def\cM{{\cal M}}
\def\cL{{\cal L}}
\def\cT{{\cal T}}
\def\surfb{S^2_n}
\def\tr{\mbox{Tr}}

\begin{document}
\title{%
Effective particle kinematics from Quantum Gravity}
\author{%
J.\ Kowalski--Glikman\thanks{Institute for Theoretical Physics,
University of Wroclaw,   Poland;  e-mail {\tt jkowalskiglikman@ift.uni.wroc.pl}},%
 $\;$ and A.\ Starodubtsev\thanks{Centre de Physique Theorique de Luminy, F-13288 Marseille}%
} \maketitle

\begin{abstract}
Particles propagating in de Sitter spacetime can be described by the
topological  BF $\SO(4,1)$ theory coupled to point charges.
Gravitational interaction between them can be introduced by adding
to the action a symmetry breaking term,  which reduces the local
gauge symmetry down to $\SO(3,1)$, and which can be treated as a
perturbation.  In this paper we focus solely on  topological
interactions which corresponds to zeroth order in this perturbative expansion.  We
show   that in this approximation the system is effectively
described by the $\SO(4,1)$ Chern-Simons theory coupled to particles
and living
 on the 3 dimensional boundary of space-time.
  Then, using Alekseev--Malkin construction we find the effective theory of  particles kinematics.
  We show that the particles action contains standard kinetic terms and the deformation shows up in the presence
   of interaction terms. The strength of the interactions is proportional to deformation parameter,
    identified with Planck mass scale.
\end{abstract}

\section{Introduction}

It is well known that in 3 dimensions gravity is described by a
topological field theory, and therefore has a finite number of
topological degrees of freedom reflecting the topology of spacetime
\cite{Witten:1988hc}. If $3d$ gravity is coupled to point particles,
which can be modeled as charged punctures of space manifold, these
topological degrees of freedom can be ``integrated out'' leading to
effective, deformed particle kinematics \cite{Matschull:1997du},
\cite{Meusburger:2005mg}, \cite{Schroers:2007ey}. This effective
system of deformed particles is of Doubly Special Relativity type
(see \cite{Amelino-Camelia:2000ge}, \cite{Amelino-Camelia:2000mn},
\cite{rbgacjkg} for the original proposal and
\cite{Kowalski-Glikman:2004qa}, \cite{Kowalski-Glikman:2006vx} for
reviews), being characterized by two scales, velocity of light $c$
and Planck mass $\kappa$. Similarly it turns out that  integrating
out topological gravitational degrees of freedom in the case of
gravity coupled to fields leads to effective field theory on
non-commutative spacetime \cite{Freidel:2005me},
\cite{Freidel:2005bb}. As a result spacetime symmetries of deformed
systems become quantum symmetries, being described by quantum
groups, instead of the standard Lie groups.

The question arises if something similar could happen in the case of
physical,  4-dimensional gravity coupled to particles and/or fields?
Certainly, in $4d$ gravity possesses local degrees of freedom
exhibited by Newtonian interactions and gravitational waves, for
example. However, as stressed in the context of recent
investigations in quantum gravity phenomenology (see
\cite{AmelinoCamelia:2008qg} for recent review and references to
earlier works) our best hope to see signals of quantum gravity is to
look for high-energetic events events (like scattering of ultra high
energy cosmic rays) in description of whose local degrees of freedom
of gravity play no role. If we expect to see some quantum gravity
imprints there, to describe them we must look for ``no local gravity limit of
gravity'', hoping that in this limit the effective theory behaves
like gravity in $3d$, effectively deforming particles and fields
kinematics. This is why we make use of DSR-like test theories, that
predict deformation of spacetime-symmetries characterized by
Planck-scale deformation parameter to describe quantum gravity
signal that may be detected in foreseeable experiments.

It is therefore of  interest to look for a formulation of gravity
such  that one keeps good control over the limit, in which local
gravitational degrees of freedom are not present. In this limit
gravity is described locally by its maximally symmetric vacuum
state: Minkowski space for zero cosmological constant, and (anti) de
Sitter space in the case of the (negative) positive one. Coupling such a
theory to point particles and then taking the limit makes it
possible investigate the effective behavior of the particles. One
then can ask the question if the quantum gravity scale is still present
in this effective theory, acting as a deformation? Naively the
answer would be in the negative, and after taking a limit we would
end up with the standard theory of particles moving on Minkowski
space of special relativity (or on (anti) de Sitter space when the
cosmological constant is non-zero.) On the other hand, the
experience of 3d gravity suggests that even in the limit the
effective theory might be deformed by the presence of Planck scale.

This is the problem that we would like to investigate in this paper.
\newline

As said above, in 4 dimensions gravity is certainly not a
topological theory.  However, it can be nevertheless described by a
BF topological theory with de Sitter $\SO(4,1)$ gauge group,
appended by a small term that breaks gauge symmetry down to Lorentz
$\SO(3,1)$. The presence of this symmetry breaking term switches on
the local degrees of freedom of gravity, while in the limit, in
which this  term  vanishes the theory becomes topological, as in
$3d$. Such formulation of gravitational field has its roots in the
proposal of MacDowell and Mansouri \cite{MacDowell:1977jt} and has
been recently investigated in depth in \cite{Smolin:2003qu} and
\cite{Freidel:2005ak}.

The construction presented in these works is as follows. The
building  blocks are the $\SO(4,1)$ connection $A^{IJ}$ with
curvature $F^{IJ}=dA^{IJ}+A^{I}{}_{K} \wedge A^{KJ}$ and the
$\so(4,1)$ valued two-form field $B^{IJ}$ ($I,J=0,\ldots,4$), in
terms of whose the action takes the form
\begin{equation}\label{0.1}
S=\int B^{IJ} \wedge F_{IJ} -\frac{\beta}{2} B^{IJ} \wedge B_{IJ}  -\frac{1}{2} B^{IJ} \wedge B^{KL} \epsilon_{IJKLM}V^{M}
\end{equation}
In this action a constant algebra element $V^M$ enforces breaking
the gauge  symmetry group down to $\SO(3,1)$ (the subgroup of
$\SO(4,1)$ that leaves $V^M$ invariant.) Without loss of generality
we can take this vector to be
$V^{M}=(0,0,0,0,V^4)=(0,0,0,0,\frac{\alpha}{2})$ so that the action
takes the form
\begin{equation}\label{0.2}
S=\int B^{IJ} \wedge F_{IJ} -\frac{\beta}{2} B^{IJ} \wedge B_{IJ}  -\frac{\alpha}{4} B^{IJ} \wedge B^{KL} \epsilon_{IJKL4}
\end{equation}
 The first two terms in the action describe the topological field theory,
 and in  the limit $\alpha\rightarrow0$ the only solution of equations of motion is
 de Sitter space, the topological vacuum of the full theory. Notice that both the topological
 lagrangian  and the gauge breaking one   are manifestly diffeomorphism invariant,
 and thus  the perturbation theory in $\alpha$, around topological vacuum,  corresponding to
  topological theory at $\alpha=0$, is going to be manifestly diffeomorphism invariant as well.

Remarkably, it can be shown \cite{Freidel:2005ak} that if one
decomposes the connection $A^{a4} = \frac1\ell\, e^a$, with $\ell$
being a constant of dimension of length,  $A^{ab}= \omega^{ab}$,
$a,b=0\ldots3$, where $e^a$ and $\omega^{ab}$ are tetrad and Lorentz connection one forms, respectively, after solving for $B$ this action reproduces, up to
topological terms, to the standard Einstein--Cartan one
\begin{equation}\label{0.3}
    S_P=-\frac{1}{2G}\int(R^{ab}\wedge e^c \wedge e^d-\frac{\Lambda}{6}e^a \wedge e^b\wedge e^c  \wedge e^d) \epsilon_{abcd}-\frac{2}{\gamma}R^{ab}\wedge e_a \wedge e_b
\end{equation}
The last term in this action does not  modify  field equations if
torsion vanishes.  The physical constants:  Newton's constant $G$,
cosmological constant $\Lambda$ and Immirzi parameter $\gamma$ are
related to the dimensional parameters of the original action
(\ref{0.2}) and the scale $\ell$ as follows
\begin{equation}\label{0.4}
    \gamma=\frac{\beta}{\alpha}, \qquad \frac{1}{\ell^2}=\frac{\Lambda}{3},
    \qquad G=\frac{3\alpha(1-\gamma^2)}{\Lambda}=\frac{3\beta(1-\gamma^2)}{\Lambda\gamma}
\end{equation}

One can couple gravity described by the action (\ref{0.2}) to
particles in a rather  straightforward way \cite{Freidel:2006hv}
(see also \cite{Fairbairn:2008hy} for recent discussion.) Each
particle with (dimensionless) mass $\mu = \ell m$ and spin $s$ at
rest is described by an appropriate element of the $\so(4,1)$
algebra\footnote{Strictly speaking it is an element of the dual
algebra $\so(4,1)^*$. Here we describe it by a canonically conjugated
element of the algebra.}
\begin{equation}\label{0.5}
   D = \mu T^{04} + s T^{23}
\end{equation}
where $T^{04}$ and $T^{23}$ are ``translational'' ``rotational''
generators of $\so(4,1)$ algebra,  respectively. Then the lagrangian
describing the particle at rest is simply
\begin{equation}\label{0.6}
   S_{rest} = \int d\tau \tr (D A_\tau), \quad A_\tau \equiv A_\mu \dot z^\mu(\tau)
\end{equation}
where $z^\mu(\tau)$ is the particle worldline, and $\tau$ is the
affine parameter.  The action (\ref{0.6}) breaks gauge invariance at
the particle worldline. However $\SO(4,1)$ gauge transformations are
just translations and Lorentz transformations, so acting by gauge
transformations on the particle at rest just makes the particle
moving. In this way the gauge degrees of freedom of gravity on
worldline become dynamical degrees of freedom of the particle and
the action describing arbitrarily moving particle has the form
\begin{equation}\label{0.7}
   S_{particle} = \int d\tau \tr (D A^h_\tau), \quad A^h_\tau \equiv h^{-1} A_\tau h + h^{-1} \partial_\tau h, \quad h \in \SO(4,1)
\end{equation}
It can be shown \cite{Freidel:2006hv} that this action leads to
correct equations of  motion for the particle and generalized
Einstein equations with mass and spin of the particle being the
source of curvature and torsion, respectively.

Notice that the action (\ref{0.7}) easily generalizes to the case of
a  finite number of particles in which case it reads
\begin{equation}\label{0.8}
   S_{particles} = \sum_i\int d\tau \tr (D_i A^h_\tau),
\end{equation}
For further discussion of this formulation of gravity coupled to
particles  see \cite{Freidel:2006hv} and
\cite{KowalskiGlikman:2006mu}.

The action of gravity (\ref{0.2}) coupled to particles (\ref{0.8})
is  a convenient starting point to address the question raised
above. The point is that the presence of the parameter $\alpha$ in
the action (\ref{0.2}) makes the ``no gravity limit'' easy to
control, and to set up a perturbative theory. In the next section we
consider the quantum perturbative expansion in parameter $\alpha$
and we  argue that the zeroth order of this expansion is described
by a holographic theory living on the boundary of spacetime. Then,
in the following two sections, we will show how such boundary theory
reduces to effective particles dynamics. In the final section we
discuss the obtained results.

\section{From gravity in $4d$ to $3d$ Chern--Simons theory }

The purpose of this section is to demonstrate that the zeroth order
approximation of perturbation theory of quantum gravity with
particles around BF topological quantum field theory is described by
Chern-Simons theory coupled to point sources. We will not study
higher order corrections here.

For shortness let us write the action (\ref{0.2}) as
\begin{equation}\label{modelgeneral}
S=\int L_{BF}+\frac\alpha4\,\int L_{I},
\end{equation}
where now $L_{BF}$ is a lagrangian of topological field   theory
possibly coupled to particles, and  $L_{I}$ is a symmetry breaking
interaction term.

The general perturbative expression for the partition function
coupled to arbitrary finite number of particles (where we neglect
all the interactions except gravitational)  looks like
\begin{eqnarray}\label{series1}
Z(\{g_{p_i}\},\{g_{p_f}\})=\int \cD A \cD B \sum\limits_n
\frac{(i\alpha)^n}{n!}\Big( \int L_{I}(x) \Big)^n \exp \left[i
\frac{1}{\beta}\int_M L_{BF}\right] \nonumber \\ =\sum\limits_n
\frac{(i\alpha)^n}{n!}Z_n(\{g_{p_i}\},\{g_{p_f}\}),
\end{eqnarray}
Here
 $g_{p_i}$ and $g_{p_f}$ are $SO(4,1)$ group elements labeling  initial and final positions and  orientations of $p$-th
 particle with respect to the selected reference point.
 They can be obtained as holonomies of connection $A$ between corresponding points
 (see \cite{KowalskiGlikman:2006mu} for detailed discussion.)

The expression (\ref{series1}) is formal, of course, and we have to
define it precisely.  First we must specify the measure $\cD A$ in
the path integral. Since the action describes a system with gauge
symmetry the conventional way to proceed would be to introduce a
gauge-fixing term. An immediate problem with this approach has been
pointed out in \cite{Cattaneo:1997eh} for Yang-Mills theory. Namely,
since the interaction term breaks the symmetry, higher order terms
in perturbative expansion have less gauge symmetry than the free
action. On the other hand, if we do not fix all the gauge symmetries
of the free action we cannot construct a propagator because of
non-invertibility of the quadratic form. The approach of
\cite{Cattaneo:1997eh} has been to introduce an  auxiliary field
that turns gauge degrees of freedom into physical ones already at
zero order. Such procedure however trivially reduces the
perturbative expansion to the standard one around  fixed background.
This  is appropriate in the case of Yang-Mills theory, but cannot be
applied in the case of a theory that is supposed to be background
independent.

In this paper we consider a different approach which does not use a
gauge fixing. To define a path integral without gauge fixing one has
to explicitly construct the reduced phase space spanned by the
complete set of gauge invariant observables and define a measure on
it. In most situations the later is not possible for technical
reasons, e.g. it is certainly impossible to construct all the
diffeomorphism invariant observables of four dimensional General
Relativity. Fortunately, this turns out to be possible for any
finite order terms of the perturbation theory considered here. This
happens because the starting point for the expansion is a
topological field theory whose reduced phase space (moduli space) is
finite dimensional. Then, in finite order of perturbative expansion
the dimensionality of the moduli space is getting larger, but
remains finite. Therefore, in any finite order of expansion the
moduli space can be,  explicitly constructed and the measure on it
can be defined.

To show this let us replace the integral over the symmetry
breaking lagrangian in (\ref{series1}) by its Riemann definition.
Divide the manifold $\cM$ into $N$ cells $\cM_i$ where each cell
is sufficiently small so that every field $\phi$ on which the
lagrangian depends can be considered constant within each cell
$$
\phi(x)\Big\vert_{\cM_i}=\phi_i
$$
The integral is a sum of contributions from every cell
$$
\int L_{Int}(\phi(x))=\sum\limits_i^N S_I(\phi_i),
$$
where
$$
S_I(\phi_i)=\int\limits_{\cM_i} L_I(\phi_i).
$$
Here we took into account that the lagrangian $L_I$ is a density
and the volume element is already contained in it.

Consider the contribution  of a cell $\cM_i$ to the partition
function at the first order $Z_{1i}=\int \cD \phi
e^{iS_T}S_I(\phi_i)$ and compare it with the contribution $Z_{1i}$
from another cell $\cM_j$. One can always find a diffeomorphism $x
\rightarrow x'$ such that $\cM_i \rightarrow \cM_j$. The fields
also transform under this diffeomorphism $\phi \rightarrow \phi'$.
Due to diffeomorpfism invariance of the interaction term
$$
S_I(\phi_i)=S_I(\phi'_j).
$$
The free action is also diffeomorphism invariant and we assume
that we can define a diffeomorphism invariant measure of the path
integral
$$
S_T(\phi)=S_T(\phi'), \ \ \ \cD \phi = \cD \phi'.
$$
As a result, the contribution to the path integral from different
cells is equal, $Z_{1i}=Z_{1j}$, and the sum over cells becomes
trivial.

At higher order one has to distinguish the situations when the
cells on which the interaction term is applied are the same or
distinct.
$$
Z_n=N^n \int\cD \phi \prod\limits_iS_I(\phi_i) e^{iS_T}_+...+N
\int\cD \phi S_I(\phi_i)^n e^{iS_T}
$$

In the limit of large number of cells the first term, where
interaction is applied to the distinct cells, will be dominating.

The above argument is analogous to that of coordinates-independence
of $n$-point functions of diffeomorphism invariant theories   (see
e.g.\ \cite{Rovelli:2005yj} and references therein).

So far we were using only diffeomorphism invariance of the free
action. Its topological invariance allows us to give a complete
definition of every term in the expansion of the the path
integral. The free action can be replaced by its discretized
version $S_T \rightarrow \sum\limits_i S_T(\phi_i)$, and the
measure can be specified as $\cD \phi \rightarrow \prod\limits_i
d\phi_i$. The resulting path integral is finite dimensional and
due to topological invariance of the free action is independent of
the discretization.

Let us now specify the model in (\ref{modelgeneral}) to be
(\ref{0.2})  with particles coupling in the $\alpha\rightarrow0$
limit, i.e.\ being defined, after integrating out B-field, by
\begin{eqnarray}\label{lbf}
L_{BF}=\frac{1}{\beta}F^{IJ}\wedge F_{IJ}+\sum\limits_i D_i^{IJ}A_{0IJ}\delta^3(x-x_i(\tau))
\end{eqnarray}
In the formula above $D_i=\mu_iT^{04}+s_iT^{23}$ is an algebra element
 defining mass and spin of $i$-th particle (both $\mu_i = m\ell$, where $m$ is
  the physical  mass and $\ell$ is the length scale of the model (\ref{0.4}),
  and $s_i$ are dimensionless here), $x_i(\tau)$ is a
  timelike particle trajectory which can be taken arbitrary due to topological invariance of the model.
In the present paper we are studying only the zeroth order contribution, and therefore
we do not consider the interaction term $L_{Int}$ here.

Due to Bianchi identity the integral of the first term in
(\ref{lbf}) after  integration by parts reduces to a Chern-Simons
action on the boundary of the original manifold $\cal{M}$.
\begin{equation}
  \frac{1}{\beta}\int_{\cal{M}}F^{AB}\wedge F_{AB}=\frac{1}{\beta}\int_{\partial {\cal{M}}}Y_{CS}(A)
\end{equation}
Below we consider the boundary $\partial \cal{M}$ to be the direct product
of a (punctured) sphere $S^2_n$ with the real line $R$.

The second term in (\ref{lbf}) breaks the gauge symmetry at the
location of the particles thus promoting some of the gauge degrees
of freedom to the physical ones. We can include the latter
explicitly by substituting into the action the connection in an
explicit gauge transformed form:
 \begin{equation}
 A \rightarrow h^{-1}d h + h^{-1}A h \label{atrans}
 \end{equation}
At the location of the particles $h(x_i)\equiv h_i$ become physical degrees of freedom.

To make a link with canonical formulation we will rewrite the
resulting action explicitly, decomposing the connection into
.spacelike and timelike components.
\begin{align}
 S_{BF}[A_S,&A_0, h_i]=
\label{2.1a}
 \int_\mathbb{R} dx^0\int_{\surfb}\,\frac{k}{4\pi}\left<\partial_0 A_S\wedge A_S\right>
 -\int_\mathbb{R} dx^0 \sum_{p=1}^n \left< D_i\,,\, h_i^{-1}\partial_0 h_i\right>\\
 &+\int_{\cal{M}} d^4x\left< A_0\,,\, \,\frac{k}{2\pi}F_S\delta(\partial {\cal{M}}) -
\sum_{p=1}^n T_i\delta^{(3)}(x-x_{(i)})dx^1\wedge dx^2 \wedge dx^3\right>
\nonumber
\end{align}

Let us explain the notation used in the formula above. The
Chern-Simon connection one-form $A$ is decomposed into time and
space part, to wit
\begin{equation}\label{2.2a}
    A = A_0\, dx^0 + A_S, \quad F = dx^0\wedge \left(\partial_0A_S- d_S A_0 +[A_0,A_S]\right) +F_S
\end{equation}
and thus $F_S$ is the space part of the curvature two-form. We  take
the Chern-Simon coupling constant to be $k/4\pi=1/\beta$.
 $D_i$ is the charge carried by the particle, which in our case
 will be just its  mass $D_i = \mu_i\, T^{04}$, while $T_i$ is the algebra element defined by
\begin{equation}\label{2.3a}
    T_i = h_iD_ih_i^{-1}
\end{equation}
Finally $<\cdot, \cdot>$ denotes the invariant, symmetric,  bilinear
form on the gauge group algebra, which below will be taken to be a
trace of product of appropriate matrices, normalized such that
$<\mathbf{1}>=1$.

The one-dimensional delta function $\delta(\partial {\cal{M}})$
reducing the expression to the boundary is defined by
\begin{equation}
\int_{\cal{M}} \delta(\partial {\cal{M}}) (\star) =\int_{\partial {\cal{M}}} (\star)
\end{equation}

In the action (\ref{2.2a}) the first line is the kinetic term,
while the second is the constraint
\begin{equation}\label{2.4a}
    \frac{k}{2\pi}F_S \delta(\partial {\cal{M}}) -
\sum_{i=1}^n T_i\delta^{(3)}(x-x_{(i)})dx^1\wedge dx^2 \wedge dx^3  =0
\end{equation}

Notice that because of topological invariance of the model the
deformation of the trajectory of a particle $x_{(i)} \rightarrow
x'_i(x_{(i)})$ does not change the value of the physical degrees of
freedom of the model. The later are encoded in the group elements
$h_i$ at the location of the particle and not in the position of the
trajectory. In particular, by such deformation one can map the whole
particle trajectory $x_{(i)}$ on the boundary $\partial {\cal{M}}$
of the manifold without changing $h_i$.

However, the constraint (\ref{2.4a}) distinguishes the boundary,
and in fact, mapping the particle trajectories to the boundary is
the only way to satisfy it for a nonzero value of $T_i$. Indeed, if
some part of the particle trajectory does not belong to the
boundary, at such points the first term in (\ref{2.4a}) is zero,
while the second term is equal to a constant times $T_i$. In such
situation the constraint (\ref{2.4a}) would force $T_i$ to be zero.
As due to Bianchi identity the charge $T_i$ has to be conserved,
$T_i=0$ along some part of trajectory means $T_i=0$ along the whole
trajectory which in turn means that there is no particle. Thus, the
only way to introduce a particle satisfying  constraint (\ref{2.4a})
is to map the whole particle trajectory on the boundary.

Mapping particle trajectories on the boundary is analogous to
introducing the Dirac string singularity for magnetic monopole in
electromagnetism. Through the Dirac string the magnetic flux can
reach from the boundary to the point at which the monopole is
located. In our model the situation is even simpler. In topological
field theory the position of a monopole with respect to a manifold
coordinates has no physical relevance. Therefore connecting a
monopole to the boundary with a string is the same as placing the
monopole on the boundary.

We have seen that the equations of motion of the model,  namely the
constraint (\ref{2.4a}), force the particle trajectories to lie on
the boundary $\partial {\cal{M}}$ of the manifold. So far our
considerations were classical. But we should expect these results to
hold also in quantum theory because the equations of motion which
are constraints on initial data, such as (\ref{2.4a}) hold in
quantum theory exactly. To see this let us substitute the action
(\ref{2.1a}) into the path integral over the connection $A$ and
perform integration over $A_0$. Among the spatial components of
connection $A$ we will distinguish $A_S$ -- the components lying
within the boundary and $A_3$ -- the component transverse to the
boundary. As the action does not depend on $A_3$ we will not include
integration over it. We obtain
\begin{eqnarray}
W_0(\{g_{p_i}\},\{g_{p_f}\})=\int \cD A e^{i S_{BF}[A_S,A_0, h_i]}= \nonumber \\
\int \cD A_S  e^{i \Big(\int_\mathbb{R} dx^0\int_{\surfb}\,\frac{k}{4\pi}\left<\partial_0 A_S\wedge A_S\right>
 -\int_\mathbb{R} dx^0 \sum_{p=1}^n \left< D_i\,,\, h_i^{-1}\partial_0 h_i\right>\Big)} \nonumber \\
\prod\limits_{x\in
\partial M }\delta\Big( \frac{k}{2\pi}F_{S,12}  -
\sum_{i=1}^n T_i\delta^{(2)}(x-x_{(i)})\Big) \nonumber \\ \prod\limits_{x \not\in
\partial M}\delta\Big(
\sum_{i=1}^n T_i\delta^{(`3)}(x-x_{(i)})\Big) \label{w0complete}
\end{eqnarray}
The last factor in the path integral in (\ref{w0complete}), $$\prod\limits_{x \not\in
\partial M}\delta\Big(
\sum_{p=1}^n T_i\delta^{(`3)}(x-x_{(i)})\Big)$$ forces  the
partition function to be zero whenever we have a particle away from
the boundary. If all the particles are on the boundary this factor
is  (an infinite) constant which can be absorbed in the
normalization of the partition function. Therefore the path integral
(\ref{w0complete}) has only contributions from the particles sitting
on the boundary where the last factor can be ignored. The resulting
path integral is a path integral for the action
\begin{align}
 S[A_S,A_0, h_i]=
\label{actionCS}
 &\int_\mathbb{R} dx^0\int_{\surfb}\,\frac{k}{4\pi}\left<\partial_0 A_S\wedge A_S\right>
 -\int_\mathbb{R} dx^0 \sum_{i=1}^n \left< D_i\,,\, h_i^{-1}\partial_0 h_i\right>\\
 +&\int_\mathbb{R} dx^0\int_{\surfb}\left< A_0\,,\, \,\frac{k}{2\pi}F_S -
\sum_{i=1}^n T_i\delta^{(2)}(x-x_{(i)})dx^1\wedge dx^2 \right>
\nonumber
\end{align}
of Chern--Simons theory on the boundary coupled to point charges.  We
will study this action in the next section.

The relation between quantum gravity in the bulk and Chern--Simons
theory on the boundary was first studied in \cite{lee95} and
extended to include translational degrees of freedom in
\cite{topexit}. For the amplitudes considered in this paper the
correspondence is precise. This picture is also supported by
spinfoam studies \cite{jb04} where it was shown that the
invariants of four dimensional Crane-Yetter model are equivalent
to Turaev-Viro invariants on the three dimensional boundary.

\section{ From Chern--Simons to particle kinematics}

Let us summarize what was achieved above. We showed that in the
topological limit the action for gravity coupled to the particles is
equivalent to the Chern-Simon action for the gauge group $\SO(4,1)$,
coupled to particles, carrying the charges (masses and spins) of the
same group. The 3 dimensional manifold on which this theory is
defined is assumed to be a product of a punctured 2-sphere $S^2_n$,
with each puncture corresponding to the particle, with real line
$\mathbb{R}$ representing time. Thus in the zeroth order of
perturbative expansion, i.e., in the ``no local gravity limit of
quantum gravity'' our theory is described by a holographic quantum
$\SO(4,1)$ Chern--Simons theory with particles.

In what follows we recall the construction of Alekseev and Malkin
\cite{Alekseev:1993rj} (see also \cite{Meusburger:2005mg}, from
which we borrowed the notation). In the case when the topology of
the boundary of spacetime is simple (no handles) and thus reflects
only the presence of particles exhibited by punctures, the
topological degrees of freedom of gravity can be absorbed by
particle's. Thus effectively we obtain a (possibly deformed) theory
of particles kinematics. As stressed above this theory could only
depend on gauge group elements, as positions in spacetime do not
play any role.

In what follows we will consider only the classical theory.  The
symplectic form on the space of gauge field and particles
configurations can be easily found from the action (\ref{actionCS})
and reads
\begin{equation}\label{2.1}
   \Omega = \frac{k}{4\pi}\int_{\surfb} \left<\delta A_S \wedge \delta A_S\right> + \sum _{i=1}^n \delta \left< D_i, h_i^{-1}\delta h_i\right>
\end{equation}
This symplectic form is subject to the constraint
\begin{equation}\label{2.2}
    \frac{k}{4\pi} F_S = \sum _{i=1}^n T_i\delta^{(2)}(x-x_{(i)})dx^1\wedge dx^2
\end{equation}
which makes the curvature zero everywhere except for  the positions
of the particles. It follows that the connection $A_S$ takes simple
form at appropriately defined submanifolds of $\surfb$.

\begin{figure}
\includegraphics[angle=0,width=10cm]{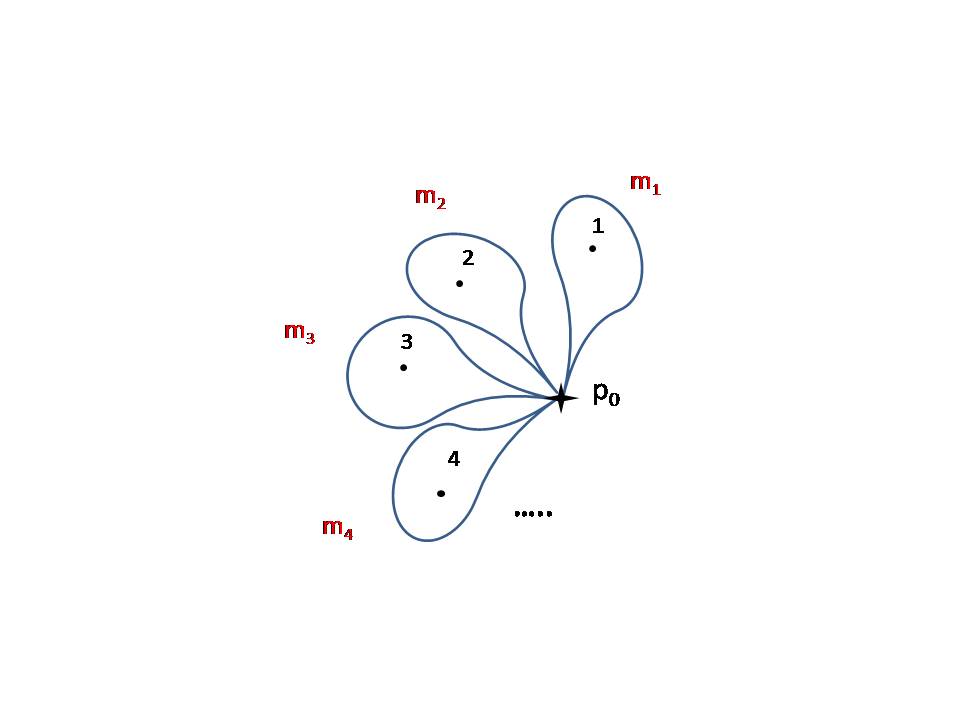}
\caption{}
\end{figure}

Alekseev and Malkin \cite{Alekseev:1993rj}  define these
submanifolds as follows. Take a point $p_0$ away from the punctures
and construct  loops $m_i$ with this base point around each puncture
(Figure 1). Along these loops we will later calculate holonomies.
Now cut the surface along the loops and remove the so obtained
discs. As a result we obtain $n$ punctured discs $Q_i$ and the
polygon $Q_0$ with no punctures inside, whose boundary will contain
exactly $n$ edges which we denote $m_i$ and vertices $p_i$ $p_0
\stackrel{m_1}{\rightarrow} p_1 \cdots p_{n-1}
\stackrel{m_n}{\rightarrow} p_0$ (Figure 2).

Now the first term in (\ref{2.1}) can be decomposed as follows
\begin{equation}\label{2.3}
 \frac{k}{4\pi}\int_{\surfb} \left<\delta A_S \wedge \delta A_S\right> =
 \frac{k}{4\pi}\int_{Q_0} \left<\delta A_S \wedge \delta A_S\right> + \frac{k}{4\pi}\sum_{i=1}^n \int_{Q_i} \left<\delta A_S \wedge \delta A_S\right>
\end{equation}

The virtue of the decomposition (\ref{2.3}) is that on each region
the form  of the connection is quite simple.  Consider the region
$Q_0$ first. Since there are no punctures in this simply connected
region the constraint (\ref{2.2}) tells that connection $A_S$ is
trivial there
\begin{equation}\label{2.4}
    A_S|_{Q_0} = \gamma_0 d_S \gamma_0^{-1}
\end{equation}
 Then by direct calculation one can convince oneself that the first
 integral in (\ref{2.3}) reduces to the boundary one
\begin{equation}\label{2.5}
  \frac{k}{4\pi}\int_{Q_0} \left<\delta A_S \wedge \delta A_S\right>  = \frac{k}{4\pi}\int_{\partial Q_0} \left<\delta \gamma_0^{-1} \gamma_0 , d(\delta \gamma_0^{-1} \gamma_0)\right>
\end{equation}

\begin{figure}
\includegraphics[angle=0,width=10cm]{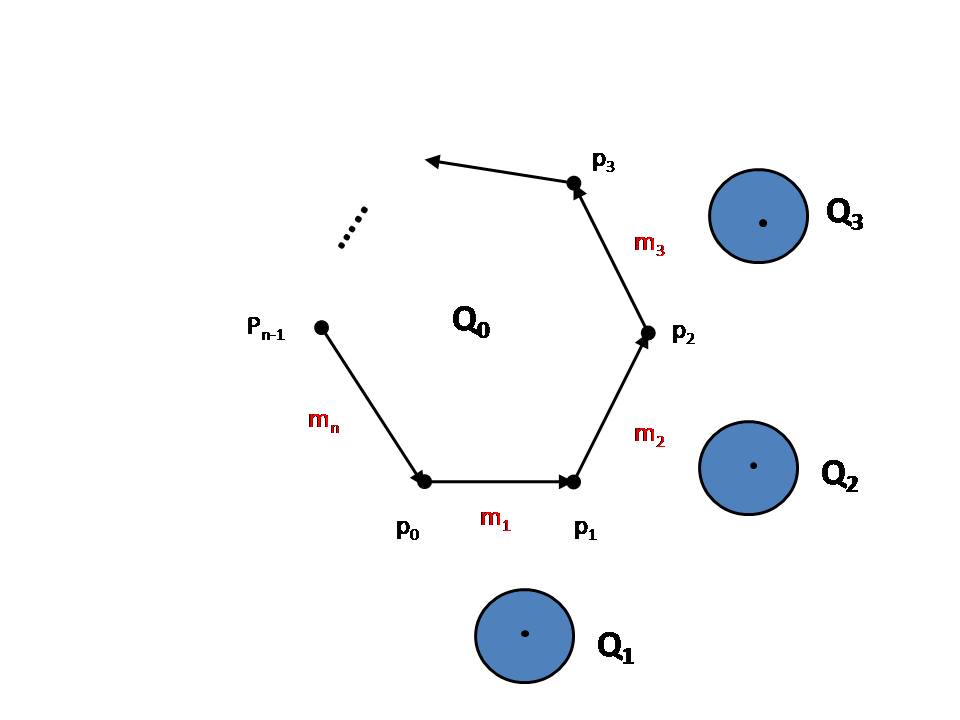}
\caption{}
\end{figure}

The contribution from the discs $Q_i$ can be found by similar
analysis.  The only difference is that now we have to do with a
region with puncture inside, carrying the charge, being an element
of gauge algebra. Since the curvature must have the delta
singularity at the puncture, the most general connections on the
disc must be given by gauge transformations of the canonical ones.
The latter are defined by
$$
B_i = \frac1k\, \tilde D_i d\phi_i, \quad \tilde{D}_i = \frac{k}{2\pi}\, \mu_i T^{04} =\frac{k}{2\pi}\, D_i
$$
where  $\phi_i$ are angular coordinates, that along with the radial
ones (defined such that the singularity corresponds to $r_i=0$) can
be introduced on the disks globally. Thus
\begin{equation}\label{2.6}
   A_S|_{Q_i}= \frac1k\,\gamma_i \tilde{D}_i d\phi_i \gamma_i^{-1} +\gamma_i d \gamma_i^{-1}
\end{equation}
It is easy to see that since $dd\phi_i=2\pi\delta(x-x_i)dx\wedge dy$
this connection solves the constraint (\ref{2.2}) if
$\gamma_i(x_i)=h_i$. Plugging (\ref{2.6}) to one of the $Q_i$
integrals in (\ref{2.3}) we find
$$
\int_{Q_i} \left<\delta A_S \wedge \delta A_S\right>
$$
\begin{equation}\label{2.7}
= - \delta\left< \tilde{D}_i, h_i^{-1}\delta h_i\right> +
\frac{k}{4\pi} \int_{\partial Q_i}\left<\delta\gamma_i^{-1}\gamma_i d\left(\delta\gamma_i^{-1}\gamma_i\right)\right>
-\frac{1}{2\pi } \int_{\partial Q_i}\delta\left<\tilde{D}_i\delta\gamma_i^{-1}\gamma_i\right> d\phi_i
\end{equation}
Notice that the sum of the first terms in (\ref{2.7}) cancels
exactly the second term in (\ref{2.1}).  Thus the symplectic form is
a sum of (\ref{2.5}) and the last two terms in (\ref{2.7}). Its form
can be simplified further by observing that the connections on the
segments of $\partial P_0 = \bigcup_i m_i$ must be equal to the
connection on appropriate $\partial P_i$, in order to make
connection continuous. We have therefore
\begin{equation}\label{2.8}
  \left.  A_S\right|_{m_i} = \left.\gamma_0 d_S \gamma_0^{-1}\right|_{m_i} =\left.\left(\frac1k\,\gamma_i \tilde{D}_i d\phi_i \gamma_i^{-1} +\gamma_i d \gamma_i^{-1}\right)\right|_{\partial Q_i} =\left.  A_S\right|_{\partial Q_i}
\end{equation}
This equation can be solved to give
\begin{equation}\label{2.9}
    \left. \gamma_0^{-1}\right|_{m_i} = N_i \exp\left(\frac1k\, \tilde{D}_i \phi_i\right)\, \left. \gamma_i^{-1}\right|_{m_i}, \quad dN_i =0
\end{equation}
Substituting this expression to (\ref{2.5}) and then,  along with
(\ref{2.7}) to (\ref{2.3}) will give us the final result.

Before doing this final step, let us consider again the polygon
$P_0$.  It has $n$ vertices $p_0, p_1, \ldots, p_{n-1}$ connected by
edges $m_i$. Since the connection is trivial on the polygon, the
parallel transport along an edge is given simply by a product of the
gauge parameters at the beginning and the end of the edge
$$
PT_{p_{i-1}\rightarrow p_i} = \gamma_0(p_i)\gamma^{-1}_0(p_{i-1})
$$
But because of the continuity of the connection the values  of
$\gamma_0$ at the vertices of the of the polygon $P_0$ are just the
products of holonomies $M_i$ of the connection of the disks
\begin{equation}\label{2.10}
    \gamma_0(p_i)\equiv K_i^{-1}=M_iM_{i-1}\cdots M_1, \quad \gamma_0(p_0)\equiv K_0^{-1} =1
\end{equation}
Next, since the connection $\gamma_i$ is single valued  on each disk
it is clear from (\ref{2.6}) that each holonomy is of the form
\begin{equation}\label{2.11}
    M_i=g_i C_i^{-1} g_i^{-1}, \quad C_i=\exp\left(\frac{2\pi}k\, \tilde{D}_i \right), \quad g_i = \gamma_i(p_i)
\end{equation}

Now we are ready to present the central result  of this section, the
Alekseev--Malkin theorem \cite{Alekseev:1993rj}: The symplectic form
$\Omega$ (\ref{2.1})  gets contributions only from vertices of the
polygon $Q_0$ and reads
\begin{equation}\label{2.12}
 \Omega=\frac{k}{4\pi}\sum_{i=1}^n\left(\left<C_ig_i^{-1}\delta g_i C_i^{-1}\wedge g_i^{-1}\delta g_i\right>-\left<\delta K_iK_i^{-1}\wedge\delta K_{i-1}K_{i-1}^{-1}\right>\right)
\end{equation}

The first term in this expression is a  deformed kinetic term (a
deformation of the standard free particle symplectic form $\sum
\delta p_\mu \wedge \delta q^\mu$, and the second describes some
``topological interaction'' of the particles.   Notice that as
expected the final form of the symplectic structure does contain no
trace of the connection; there is as many degrees of freedom as the
number of particles, each described by a gauge group element. In
this way the topological degrees of freedom of gravity has been
``eaten'' by the particle's ones.

We will discuss the theory of particles  provided by this symplectic
structure in the next section.

\section{The deformed particle}

Having obtained the symplectic form for single  particle
(\ref{2.12}), let us try to understand in which sense it describes a
deformed particle. To do that let us first compare it with with an
analogous expression for free particle moving in 4 dimensional de
Sitter spacetime in the limit of vanishing cosmological constant.

The Lagrangian for such particle is given by equation (\ref{0.7})
and reads (following \cite{Freidel:2006hv} we represent $\SO(4,1)$
generators by gamma matrices, $T^{IJ} = \gamma^{IJ} \equiv 1/2
[\gamma^I, \gamma^J]$, $I,J= 0, \ldots, 4$ with $\gamma^5$ matrix
denoted here $\gamma^4$)
\begin{equation}\label{3.1}
    L=<D A^h_\tau> = \left<\mu \gamma^{04} (h^{-1} A_\tau^{(0)} h + h^{-1}\dot h)\right>, \quad
\end{equation}
where $\mu = \ell m$ with $\ell$ being the cosmological length scale, related to the cosmological constant, cf.\ (\ref{0.4}).
Since de Sitter background is gauge equivalent to zero configuration we take $A_\tau^{(0)}=0$
Then (\ref{3.1})  takes the form
\begin{equation}\label{3.2}
    L=  \left<  \mu \gamma^{04}h^{-1}\dot h\right>
\end{equation}
Let us decompose the $\SO(4,1)$ group element into translational and Lorentz $\SO(3,1)$ parts (using Cartan decomposition)
\begin{equation}\label{3.2a}
h ={\cal T\, L}, \quad {\cal T}= \mathbf{1}+\frac{q_a}\ell\, \gamma^{a4} + O\left(\frac1{\ell^2}\right),
\end{equation}
Substituting this into (\ref{3.2}) and keeping only the  leading
term after simple calculation we find (the $\cL^{-1}\dot \cL$ term
is proportional to $\gamma^{ab}$ and cancels under the trace)
\begin{equation}\label{3.3}
    L=  \left<\frac\mu\ell\,\gamma^{04} \cL^{-1} \gamma^{a4} \cL \dot q_a\right>, \quad a=0, \ldots, 3
\end{equation}
After some $\gamma$ matrices algebra this expression gives
\begin{equation}\label{3.4}
   L =  p_a \dot q^a, \quad p_a = m \left<\gamma^{0} \cL^{-1} \gamma_{a} \cL\right>
\end{equation}
and it follows that $p_a$ defined above is to be identified with particle momentum.

Notice that it follows from (\ref{3.4}) that
\begin{equation}\label{3.4a}
   p_a\gamma^a = m  \cL \gamma_{0} \cL^{-1}
\end{equation}
This equation just says that there is one to one  correspondence
between momentum of a particle and the Lorentz transformation that
boosts the particle from the rest to its actual velocity. It follows
that the components of momenta on the left hand side are restricted
to be on shell, $p^2 +m^2 =0$. As usual we can treat these
components as independent adding to the lagrangian the on shell
constraint $p^2 +m^2$. Thus we conclude that the lagrangian
(\ref{3.1}) describes a free relativistic particle, as it should.

Returning to the starting point, eq.\ (\ref{3.1})  it can be easily
computed that the Lagrangian $L=<D h^{-1} \dot h>$ corresponds to
the symplectic form
$$
\Omega_{free} = <D \delta h^{-1} \wedge \delta h> =  -\frac12 \left< [ D, h^{-1}\delta h] \wedge h^{-1}\delta h\right>
$$
Comparing this with the first term in (\ref{2.12})  we see that as
the result of deformation instead of commutator with algebra element
$D=\mu \gamma^{04}$, in the deformed case we have to do with
conjugation with group element
 $$C=e^{2\pi D/k}=\exp(2\pi\mu \gamma^{04}/k) = \cosh\frac{2\pi\mu}k\, \mathbf{1} + \sinh\frac{2\pi\mu}k\,\gamma^{04}$$

  It can be checked that the first ``free'' term in (\ref{2.12}) reduces to
\begin{equation}\label{3.5}
\Omega=\frac{k}{4\pi}\cosh\frac{2\pi\mu}k\,\sinh\frac{2\pi\mu}k\,\left<[\gamma^{04},\,g^{-1}\delta g] \wedge g^{-1}\delta g\right>
\end{equation}

Now we have to recall that the Chern--Simons coupling  constant is
related to the original coupling constant $\beta$: $k/4\pi = 1/\beta
\sim \ell^2\kappa^2$, where $\kappa = \hbar/G$. Therefore the
prefactor in $\Omega$ in (\ref{3.5}) becomes, in the leading order
in $\ell$, $\frac12 \, m\ell$. One can then easily check using
definition of momenta (\ref{3.4a}) and the expansion (\ref{3.2a})
that (\ref{3.5}) exactly reproduces (up to the sign and a prefactor
depending on Immirzi parameter) the free particle symplectic
structure. We conclude that the first term in (\ref{2.12}) is just a
sum of free particle actions.

Let us now turn to the second, interaction term, in the symplectic form (\ref{2.12}).
As an example consider the case of two particles. In this case the interaction term  reads
\begin{equation}\label{3.6}
-\frac{k}{4\pi}\left<\delta K_2 K_2^{-1}\wedge \delta K_1 K_1^{-1}\right>, \quad K_1 = M_1^{-1}, \quad K_2 = M_1^{-1}\, M_2^{-1}
\end{equation}
To calculate this let us first expand the holonomy in  powers of
$\ell$, by making use of the Cartan decomposition of the group
(\ref{3.2a}) and definition of momenta (\ref{3.4a})
\begin{equation}\label{3.7}
   M^{-1}= g C g^{-1}= \mathbf{1} + \frac1{\kappa^2\ell}\, p_a\gamma^a\gamma^4  +O\left(\frac1{\ell^2}\right)
\end{equation}
Thus
$$
-\frac{k}{4\pi}\left<\delta K_2 K_2^{-1}\wedge \delta K_1 K_1^{-1}\right> = $$
\begin{equation}\label{3.8}
-\frac1{\kappa^2}\left(\delta p^{(2)}{}_a +\delta p^{(1)}{}_a\right) \wedge \delta p^{(1)}{}_b \left<\gamma^a\gamma^4\gamma^b\gamma^4\right> =\frac1{\kappa^2}\, \delta p^{(2)}{}_a \wedge \delta p^{(1)}{}^a
\end{equation}
Equation (\ref{3.8}) can be easily generalized to an  arbitrary
number of particles, and knowing the symplectic form one can readily
reproduce the lagrangian for the n-particles system. It is a sum of
the standard kinetic terms and the on shell constraints for each
particle along with the interaction terms
\begin{equation}\label{3.9}
   L = \sum_{i=1}^n \left(p^{(i)} \cdot \frac{d}{d\tau}\, q^{(i)} +\lambda^{(i)}\left[ p^{(i)}{}^2 + m^{(i)}{}^2\right] + \frac1{\kappa^2} p^{(i)}\cdot \sum_{j=1}^{i-1} \frac{d}{d\tau}\, p^{(j)} \right)
\end{equation}
where $\tau$ is the affine parameter and $\lambda(\tau)$  is the
Lagrange multiplier enforcing the mass shell constraints, and
$\cdot$ is the standard Minkowski product.

Equation (\ref{3.9}) is the final result of our paper.  It shows
that in the action of $n$-particles system, the actions of each
particle is not deformed while the deformation arises in the form of
the presence of the additional interaction terms. It is worth
noticing that this deformation does not change equations of motion
(because the momenta $p^{(i)}$ are constants of motion), while it
certainly changes the Poisson brackets. In particular the brackets
of positions will not vanish anymore. However, it is easy to see
that by simple change of variables
\begin{equation}\label{3.10}
    q^{(i)} \longrightarrow \tilde q^{(i)} = q^{(i)}+\frac1{\kappa^2} \sum_{j=1}^{i-1} \frac{d}{d\tau}\, p^{(j)}
\end{equation}
the lagrangian (\ref{3.9}) can be cast into the form of  the one of
the standard system of free relativistic particles.

\section{Discussion}

The results of the preceding section can be regarded surprising.
Building on the experience with $3d$ gravity one would expect that
the deformation of the final particle action is essentially
guaranteed, and the  recent investigations reported in
\cite{Meusburger:2008dc} suggested that the deformation was to be of
$\kappa$-Poincar\`e form \cite{Lukierski:1992dt}. Instead what we
got was just a standard undeformed relativistic particles action.
Let us therefore try to understand this result.

To get some more insight let us try to investigate how the  results
above change if we go beyond the the leading order in large
cosmological scale $\ell$ expansion. Our starting point is again the
formula (\ref{3.5})
\begin{equation}\label{4.0}
\Omega=\frac{k}{4\pi}\,\sinh\frac{4\pi\mu}k\,\left<\gamma^{04}\,g^{-1}\delta g \wedge g^{-1}\delta g\right>
\end{equation}
Using the definition of momenta and Cartan decomposition as  above
one computes
\begin{equation}\label{4.0a}
\Omega=\frac{k}{4\pi m}\,\sinh\frac{4\pi\mu}k\left(p_a\left<\gamma^{a4}\,\cT^{-1}\delta \cT \wedge \cT^{-1}\delta \cT\right>-\delta p_a \wedge \left<\gamma^{a4}\,  \cT^{-1}\delta \cT\right>\right)
\end{equation}
with momenta restricted to be on-shell $p^2+m^2=0$. It is  now
convenient to parametrize the translational part of the group as
follows
\begin{equation}\label{4.1}
    \cT = Q_4\, \mathbf{1} + Q_a\, \gamma^a\gamma^4
\end{equation}
with (dimensionless) positions $Q$ belonging to de Sitter space
\begin{equation}\label{4.2}
    Q_4^2+ Q^2=1,\quad Q^2 \equiv - Q_0^2 + \vec{Q}{}^2
\end{equation}
Plugging this into the symplectic form above after  straightforward
calculations we find
\begin{equation}\label{4.3}
    \Omega=\frac{k}{4\pi m}\,\sinh\frac{4\pi\mu}k\, \delta\left(Q_4\, p_a\, \delta Q^a + \frac1{Q_4}\, p_aQ^a\, Q_b\, \delta Q^b\right)
\end{equation}
It follows that the lagrangian is again up to the  irrelevant
prefactor just the one of a relativistic particle moving on de
Sitter background: defining $q^a = \ell Q^a$ we have
\begin{equation}\label{4.4}
    L = \frac1\ell\, \sqrt{1 - q^2/\ell^2}\, p_a\dot q^a + \frac1{\ell^3}\,\frac1{\sqrt{1 - q^2/\ell^2}}\, p_aq^a\, q_b\dot q^b + \lambda\left[p^2+m^2\right]
\end{equation}
Notice that again we see no trace of any deformation  and both the
symplectic structure and the lagrangian are linear in momenta, which
makes the position space commutative.

However we were not able yet to calculate the  second
``interaction'' terms of the symplectic structure beyond the leading
term in large $\ell$ expansion, which may change the picture
considerably. It is a general result of Alekseev and Malkin \cite{Alekseev:1993rj} that
such interactions can be removed by an appropriate symplectic
transformation. In this transformation the Borel subroup of the
original gauge group is known to play a role, and on the other hand
it is directly related to $\kappa$-Poincar\`e algebra and
$\kappa$-Minkowski space \cite{Freidel:2007hk}. One should also
remember that in order to get the $\kappa$-Poincar\`e algebra as a
contraction limit of $\ell\rightarrow\infty$, a nontrivial rescaling
of momenta is required (see \cite{Lukierski:1992dt} and
\cite{AmelinoCamelia:2003xp} for details.) Although such rescaling
is well motivated mathematically, it is not clear how to justify it
physically. We are going to address all these questions in the
forthcoming paper.

\section*{Acknowledgment} We would like to  thank Florian Girelli,  Catherine Meusburger,
and especially Bernd Schroers for many discussions. Thanks are also due
to Lee Smolin for reading the preliminary version of the manuscript and his constant encouragement.
For JKG this research was supported as a part of 2007-2010 research project N202 081
32/1844. AS is grateful to ENRAGE network for the support at
Utrecht University where this work started.


\begin{thebibliography}{99.}

\bibitem{Witten:1988hc}
  E.~Witten,
  ``(2+1)-Dimensional Gravity as an Exactly Soluble System,''
  Nucl.\ Phys.\  B {\bf 311} (1988) 46.

\bibitem{Matschull:1997du}
  H.~J.~Matschull and M.~Welling,
  ``Quantum mechanics of a point particle in 2+1 dimensional gravity,''
  Class.\ Quant.\ Grav.\  {\bf 15}, 2981 (1998)
  [arXiv:gr-qc/9708054].

\bibitem{Schroers:2007ey}
  B.~J.~Schroers,
  ``Lessons from (2+1)-dimensional quantum gravity,''
  arXiv:0710.5844 [gr-qc].

\bibitem{Meusburger:2005mg}
  C.~Meusburger and B.~J.~Schroers,
  ``Phase space structure of Chern-Simons theory with a non-standard
  puncture,''
  Nucl.\ Phys.\  B {\bf 738}, 425 (2006)
  [arXiv:hep-th/0505143].

\bibitem{Amelino-Camelia:2000ge}
G.~Amelino-Camelia, ``Testable scenario for relativity with
minimum-length,'' Phys.\ Lett.\ B {\bf 510}, 255 (2001)
[arXiv:hep-th/0012238].

\bibitem{Amelino-Camelia:2000mn}
G.~Amelino-Camelia, ``Relativity in space-times with short-distance
structure governed by an observer-independent (Planckian) length
scale,'' Int.\ J.\ Mod.\ Phys.\ D {\bf 11}, 35 (2002)
[arXiv:gr-qc/0012051].
\bibitem{jkgminl} J.~Kowalski-Glikman,
``Observer independent quantum of mass,'' Phys.\ Lett.\ A {\bf 286}
(2001) 391 [arXiv:hep-th/0102098].

\bibitem{rbgacjkg} N.~R.~Bruno, G.~Amelino-Camelia and J.~Kowalski-Glikman,
``Deformed boost transformations that saturate at the Planck
scale,'' Phys.\ Lett.\ B {\bf 522} (2001) 133
[arXiv:hep-th/0107039].

\bibitem{Kowalski-Glikman:2004qa}
  J.~Kowalski-Glikman,
  ``Introduction to doubly special relativity,''
  Lect.\ Notes Phys.\  {\bf 669} (2005) 131
  [arXiv:hep-th/0405273].

\bibitem{Kowalski-Glikman:2006vx}
  J.~Kowalski-Glikman,
  ``Doubly special relativity: Facts and prospects,''
  arXiv:gr-qc/0603022.

\bibitem{Freidel:2005me}
  L.~Freidel and E.~R.~Livine,
  ``Effective 3d quantum gravity and non-commutative quantum field theory,''
  Phys.\ Rev.\ Lett.\  {\bf 96}, 221301 (2006)
  [arXiv:hep-th/0512113].

\bibitem{Freidel:2005bb}
  L.~Freidel and E.~R.~Livine,
  ``Ponzano-Regge model revisited. III: Feynman diagrams and effective  field
  theory,''
  Class.\ Quant.\ Grav.\  {\bf 23}, 2021 (2006)
  [arXiv:hep-th/0502106].

\bibitem{AmelinoCamelia:2008qg}
  G.~Amelino-Camelia,
  ``Quantum Gravity Phenomenology,''
  arXiv:0806.0339 [gr-qc].











\bibitem{MacDowell:1977jt}
  S.~W.~MacDowell and F.~Mansouri,
  ``Unified Geometric Theory Of Gravity And Supergravity,''
  Phys.\ Rev.\ Lett.\  {\bf 38} (1977) 739
  [Erratum-ibid.\  {\bf 38} (1977) 1376].

\bibitem{Smolin:2003qu}
  L.~Smolin and A.~Starodubtsev,
  ``General relativity with a topological phase: An action principle,''
  arXiv:hep-th/0311163.

\bibitem{Freidel:2005ak}
  L.~Freidel and A.~Starodubtsev,
  ``Quantum gravity in terms of topological observables,''
  arXiv:hep-th/0501191.

\bibitem{Freidel:2006hv}
  L.~Freidel, J.~Kowalski-Glikman and A.~Starodubtsev,
  ``Particles as Wilson lines of gravitational field,''
  Phys.\ Rev.\  D {\bf 74} (2006) 084002
  [arXiv:gr-qc/0607014].

\bibitem{Fairbairn:2008hy}
  W.~J.~Fairbairn,
  ``On gravitational defects, particles and strings,''
  arXiv:0807.3188 [gr-qc].

\bibitem{KowalskiGlikman:2006mu}
  J.~Kowalski-Glikman and A.~Starodubtsev,
  ``Can we see gravitational collapse in (quantum) gravity perturbation
  theory?,''
  arXiv:gr-qc/0612093.

\bibitem{Cattaneo:1997eh}
  A.~S.~Cattaneo, P.~Cotta-Ramusino, F.~Fucito, M.~Martellini, M.~Rinaldi, A.~Tanzini and M.~Zeni,
  ``Four-dimensional Yang-Mills theory as a deformation of topological BF
  theory,''
  Commun.\ Math.\ Phys.\  {\bf 197} (1998) 571
  [arXiv:hep-th/9705123].

\bibitem{Rovelli:2005yj}
  C.~Rovelli,
  ``Graviton propagator from background-independent quantum gravity,''
  Phys.\ Rev.\ Lett.\  {\bf 97}, 151301 (2006)
  [arXiv:gr-qc/0508124].


\bibitem{lee95}
Lee Smolin,
Linking topological quantum field theory and nonperturbative quantum gravity.
J.Math.Phys.36:6417-6455,1995.
e-Print: gr-qc/9505028

\bibitem{topexit}
Artem Starodubtsev,
 Topological excitations around the vacuum of quantum gravity. 1. The Symmetries of the vacuum.
e-Print: hep-th/0306135

\bibitem{jb04}
  J.~W.~Barrett, J.~M.~Garcia-Islas and J.~F.~Martins,
  ``Observables in the Turaev-Viro and Crane-Yetter models,''
  J.\ Math.\ Phys.\  {\bf 48}, 093508 (2007)
  [arXiv:math/0411281].


\bibitem{Alekseev:1993rj}
  A.~Y.~Alekseev and A.~Z.~Malkin,
  ``Symplectic structure of the moduli space of flat connection on a Riemann
  surface,''
  Commun.\ Math.\ Phys.\  {\bf 169}, 99 (1995)
  [arXiv:hep-th/9312004].

\bibitem{Meusburger:2003ta}
  C.~Meusburger and B.~J.~Schroers,
  ``Poisson structure and symmetry in the Chern-Simons formulation of
  (2+1)-dimensional gravity,''
  Class.\ Quant.\ Grav.\  {\bf 20} (2003) 2193
  [arXiv:gr-qc/0301108].

\bibitem{Meusburger:2008dc}
  C.~Meusburger and B.~J.~Schroers,
  ``Generalised Chern-Simons actions for 3d gravity and kappa-Poincare
  symmetry,''
  arXiv:0805.3318 [gr-qc].

\bibitem{Lukierski:1992dt}
  J.~Lukierski, A.~Nowicki and H.~Ruegg,
  ``New quantum Poincare algebra and k deformed field theory,''
  Phys.\ Lett.\  B {\bf 293} (1992) 344.



\bibitem{Freidel:2007hk}
  L.~Freidel, J.~Kowalski-Glikman and S.~Nowak,
  ``Field theory on $\kappa$--Minkowski space revisited: Noether charges and
  breaking of Lorentz symmetry,''
  Int.\ J.\ Mod.\ Phys.\  A {\bf 23} (2008) 2687
  arXiv:0706.3658 [hep-th].

\bibitem{AmelinoCamelia:2003xp}
  G.~Amelino-Camelia, L.~Smolin and A.~Starodubtsev,
  ``Quantum symmetry, the cosmological constant and Planck scale
  phenomenology,''
  Class.\ Quant.\ Grav.\  {\bf 21} (2004) 3095
  [arXiv:hep-th/0306134].


\end{thebibliography}
\end{document}